# Tight binding approximation for bilayer graphene and nanotube structures: from commensurability to incommensurability between the layers

D.V. Chalin[1], D. I. Levshov[1,2], A.E. Myasnikova[1], S.B. Rochal[1]

[1]*Faculty of Physics, Southern Federal University, 5 Zorge str., 344090, Rostov-on-Don, Russia*
[2]*Department of Physics, University of Antwerp, B-2610 Antwerpen, Belgium*

One- and two-dimensional twisted bilayer structures are examples of ultra-tunable quantum materials that are considered as the basis for the new generation of electronic and photonic devices. Here we develop a general theory of the electron band structure for such commensurate and incommensurate bilayer graphene structures within the framework of the tight binding approximation. To model the band structure of commensurate twisted bilayer graphene (TBLG), we apply the classic zone folding theory. The latter leads us to the construction of TBLG Hamiltonians in the basis of shifted Bloch wave functions (SBWF), which, in contrast to the usual Bloch functions, have the wave vector $q$ shifted by a set of vectors $Q_i$. The dimension of the considered Hamiltonians is equal to $4T$, where the factor $T$ is a number of vertices $Q_i$ of the folded reciprocal space falling into the original first Brillouin zone of any of the layers. We propose and discuss a method for choosing a reduced set of SBWFs to construct effective Hamiltonians that correctly describe the low-energy spectrum of commensurate TBLG. The flattening of low-energy bands with a decrease in twist angle is discussed. As we show, this spectrum results from interactions between the lowest-energy modes of the folded dispersion curves. The effective Hamiltonians for calculating the low-energy band structure of incommensurate TBLG and double-walled carbon nanotubes (DWCNTs) are constructed in a similar way. To test the developed theory, we calculate the energies of 105 intra-tube optical transitions in 29 DWCNTs and compare them with previously published and novel experimental data. We also apply the theory to calculate the energies of recently discovered inter-tube transitions. Geometrical conditions allowing this type of transitions are discussed. We show that these transitions occur in DWCNTs which layers have close chiral angles and the same handedness, or in the structural context, in DWCNTs with a large unit cell of the periodic moiré pattern.

## I. INTRODUCTION

Monolayer graphene is an extraordinary material in which electron dispersion obeys a linear law and electrons being massless Dirac fermions propagate with the constant Fermi velocity [1]. Although the translational symmetry is preserved when two layers of graphene are stacked as in graphite, the characteristic masslessness of electrons disappears, as the dispersion becomes quadratic [2,3]. In turn, the relative rotation of the graphene layers breaks the translational symmetry, and we obtain another unique material, so-called twisted bilayer graphene (TBLG), the physical properties of which radically depend on the twist angle $\theta$ [4–7]. For example, additional van Hove singularities appear in the electron density of states (EDoS) of TBLG [8]. At certain "magic angles" (the largest of which is about $1.1°$ [9,10]), the band structure of TBLG rearranges in a critical way, namely, the dispersion curves in the spectrum become flat, and the Fermi velocity near the Dirac points vanishes [8–10]. The latter leads to the significant increase in the EDoS, which in combination with the specific interlayer coupling allows for superconducting and correlated insulating states in TBLG, which have been extensively studied in more than a dozen theoretical and experimental works [9–14].

At certain twist angles $\theta$ the top and bottom layers have a common subgroup of translations, and TBLG has a periodic structure, the period of which either coincides with or is expressed in multiples of periods of the emerging moiré pattern (MP) [15]. This fact simplifies the calculation of the band structure within the nearest-neighbor tight binding approximation (NN TBA) [16,17], however, at small angles $\theta$ the periodicity of the superlattice unit cell significantly increases, and one has to solve the problem of eigenvalues for matrices of very large dimensions [18]. Many theoretical approaches used for the study of incommensurate TBLG are based on the selection of an approximate unit cell [15,19], the period of which coincides with the one of MP. Such an approach in combination with different tight binding continuum models has been used on multiple occasions to calculate the low-energy region of the band structure of incommensurate TBLG [15,20–23]. Within the framework of a similar theory, the first magic angles corresponding to the band flattening were also calculated [24].

In this work we propose a different unified approach that allows one to calculate the band structure of both commensurate and incommensurate bilayer carbon nanostructures. We start from the commensurate case. Compared to the non-twisted bilayer graphene, the area of the TBLG unit cell increases by a certain number of times $T$, while the area of the first Brillouin zone (FBZ) of the reciprocal space is reduced by the same number. Therefore, the forming of the band structure of such a bilayer system can be considered within the concept of zone folding [25]: four electronic bands of two graphene sheets are folded into a new reduced FBZ of TBLG and modified due to interlayer coupling. Considering this folding, we construct the Hamiltonian in the basis of shifted Bloch wave functions (SBWFs). The SBWF differs from the ordinary Bloch function in that its wave vector $q$ is shifted by some constant reciprocal space vector $Q_i$ of the adjacent layer. Constructing a Hamiltonian of commensurate TBLG, we use a basis of $4T$ linearly independent SBWFs with shift vectors coinciding with the nodes of the reduced reciprocal lattice, which fall into the FBZ of a graphene monolayer. The proposed Hamiltonian has a simple quasi-diagonal block structure. When considering only the low-energy spectrum, the Hamiltonian can be greatly simplified by retaining only the modes related to those Dirac valleys of the adjacent layers, which become translationally equivalent after the zone folding. Moving on to the incommensurate bilayer graphene, we also apply a similar approach and construct the effective Hamiltonian allowing to calculate the low-energy region of the band structure. This Hamiltonian contains only blocks corresponding to pairwise interactions of states in the vicinity of three FBZ vertices from the first and second graphene sheets. Accordingly, the basis of the effective

Hamiltonian includes only critical SBWFs for which the quantity $\boldsymbol{q} - \boldsymbol{Q}_i$ is close to the origin of the reciprocal space.

We generalize the developed theory and apply it to analyze the band structure of incommensurate double-walled carbon nanotubes (DWCNTs). In these objects, as in TBLG, strong interlayer coupling in combination with the geometric features can also lead to interesting changes in the electron spectrum. In particular, a novel rearrangement of the band structure was demonstrated [26–28]. It allows for electron transitions between bands originating from different layers of DWCNT. Accordingly, additional peaks, which cannot be associated with the intra-tube transitions, emerge in DWCNT optical spectra. So far, the effect has been experimentally observed only in the DWCNT (17,16)@(12,11) [26] and theoretically considered for DWCNTs (15,13)@(21,17), (12,12)@(21,13), (10,6)@(14,13) [27,28]. Along with the theory of effective Hamiltonians, we develop a simple method for the approximate calculation of inter-layer coupling matrix elements in incommensurate structures and apply the method to calculate the energies of optical transitions in 30 DWCNTs, including those where we believe the recently discovered inter-tube transitions can occur [27,28]. As we show, such transitions occur in DWCNTs with the large MP unit cell and, therefore, with the same handedness and close chiral angles of the inner and outer layers.

The rest of the paper is organized as follows. The next section is devoted to the theory of the band structure in commensurate TBLG within the concept of zone folding. The SBWFs are also introduced in this section. In the third section we develop the theory of effective Hamiltonians for incommensurate DWCNTs and incommensurate TBLG. In the fourth section we compare our theory with experimental data on the optical spectra of DWCNTs. We also consider the DWCNTs, where inter-tube transitions are possible. The final section of the work is devoted to discussion of the results obtained. In addition, in this section, we consider the geometrical criteria for selecting DWCNTs, in which inter-tube electron transitions are possible.

## II. COMMENSURATE TWISTED BILAYER GRAPHENE

Within the nearest-neighbor tight binding approximation [16,17,29] the Hamiltonian of a graphene monolayer can be written as

$$H = \begin{pmatrix} 0 & f(\boldsymbol{q}) \\ f^*(\boldsymbol{q}) & 0 \end{pmatrix}, \quad (1)$$

where the quantity $f(\boldsymbol{q}) = \gamma\{\exp[-I\boldsymbol{q} \cdot (\boldsymbol{a}_1 + \boldsymbol{a}_2)/3] + \exp[I\boldsymbol{q} \cdot (2\boldsymbol{a}_1 - \boldsymbol{a}_2)/3] + \exp[I\boldsymbol{q} \cdot (2\boldsymbol{a}_2 - \boldsymbol{a}_1)/3]\}$ is the matrix element describing the interaction between the grapheme sublattices $A$ and $B$, $\gamma$ is the hopping coefficient, $\boldsymbol{a}_1$ and $\boldsymbol{a}_2$ are the graphene basis translations, $\boldsymbol{q}$ is two-dimensional wave vector, $I$ is imaginary unit. Eigen energies of the Hamiltonian (1) read $E^\pm = \pm|f(\boldsymbol{q})|$, while eigen vectors can be found as

$$|\psi^\pm\rangle = |\psi_A\rangle \pm e^{-I\varphi}|\psi_B\rangle, \quad (2)$$

where $|\psi_A\rangle$ and $|\psi_B\rangle$ are the Bloch wave functions (BWFs) of the sublattices [1]. In the coordinate representation these functions are written as:

$$\psi_\alpha(\boldsymbol{r}) = \frac{1}{\sqrt{N}} \sum_{\boldsymbol{R}_\alpha} \phi(\boldsymbol{r} - \boldsymbol{R}_\alpha) e^{I\boldsymbol{q} \cdot \boldsymbol{R}_\alpha}, \quad (3)$$

where $\phi(\boldsymbol{r} - \boldsymbol{R}_\alpha)$ is the atomic $p_z$-orbital localized near the site $\boldsymbol{R}_\alpha$, $N$ is the number of atoms in the $\alpha$ sublattice, $\alpha = A, B$. In Eq. (3) the sum goes over all the nodes $\boldsymbol{R}_\alpha$. Positive and negative signs in Eq. (2) correspond to states in the conduction band (CB) and valence band (VB) respectively, $\varphi = \text{Arg}(f/|f|)$ is the phase shift between the sublattices. Thus, the electronic states in a monolayer graphene can be classified using the vector $\boldsymbol{q}$ and the number $\sigma = \pm 1$, indexing the conduction (+1) and valence (-1) bands. Within the framework proposed below, one can calculate the band structure of a commensurate twisted bilayer graphene without additional assumptions, however, evaluating the Hamiltonian eigen energies for a bilayer with a large unit cell can be computationally expensive.

Let us consider two graphene layers stacked in $AA$ packing [15] and then let us rotate one of the layers by an angle $\theta$. At certain values of $\theta$ both layers acquire the common translation subgroup, which determines the periodicity of the bilayer structure. The following two minimal translations are usually chosen as a basis [15]:

$$\boldsymbol{C}_1 = h\boldsymbol{a}_1 + k\boldsymbol{a}_2,$$
$$\boldsymbol{C}_2 = (h+k)\boldsymbol{a}_1 - k\boldsymbol{a}_2,$$

The vector $\boldsymbol{C}_1$ is known as the chirality vector, and its components $(h, k)$, which are coprime numbers, are called the chirality indices [29]. Chirality vector $(h, k)$ corresponds to the following twist angle $\theta_{hk} = \arccos[(h^2 + 4hk + k^2)/(2h^2 + 2hk + 2k^2)]$ and triangulation factor $T = h^2 + hk + k^2$. Note that in the case when $h$ and $k$ difference is a multiple of 3, the structure $(h, k)$ has translations that are $\sqrt{3}$ times shorter than $\boldsymbol{C}_1$ (or $\boldsymbol{C}_2$). In other words, such a structure turns out to be equivalent to the one with the chirality vector $(h', k')$, where $h' = (2k + h)/3$ and $k' = (h - k)/3$.

Figure 1 (panels a and b) shows the first two commensurate bilayer structures with $h = 2, k = 1, T = 7$ and $h = 3, k = 1, T = 13$. A simple geometric analysis shows that in a twisted bilayer graphene with a factor $T$, there are $4T$ carbon atoms per primitive hexagonal cell equally divided between the top and bottom layers. The reciprocal space of both layers (see Fig. 1, panels c and d) is folded $T$ times so that a common hexagonal reciprocal space is formed, which is divided into Brillouin zones with $T$ times smaller area. If we use the scheme of extended zones [25,27], then exactly $T$ nodes (which are centers of honeycombs with coordinates $\boldsymbol{Q}_j$) of the hexagonal lattice of the common reciprocal space will fall inside the first Brillouin zone of each of the layers. Consequently, we can characterize the electron states in commensurate TBLG by the following $4T$ shifted Bloch wave functions:

$$\psi_\zeta^j(\boldsymbol{r}) = \frac{1}{\sqrt{N}} \sum_{\boldsymbol{R}_\zeta} \phi(\boldsymbol{r} - \boldsymbol{R}_\zeta) e^{I(\boldsymbol{q} - \boldsymbol{Q}_j) \cdot \boldsymbol{R}_\zeta}, \quad (4)$$

where the index $\zeta$ goes over the sublattices $A, B$ and $A', B'$ of top and bottom layers respectively and $j = 1, 2 \ldots T$. Accordingly, the calculation of the TBLG band structure reduces to the eigenvalues problem for the Hamiltonian with $4T$ dimensions.

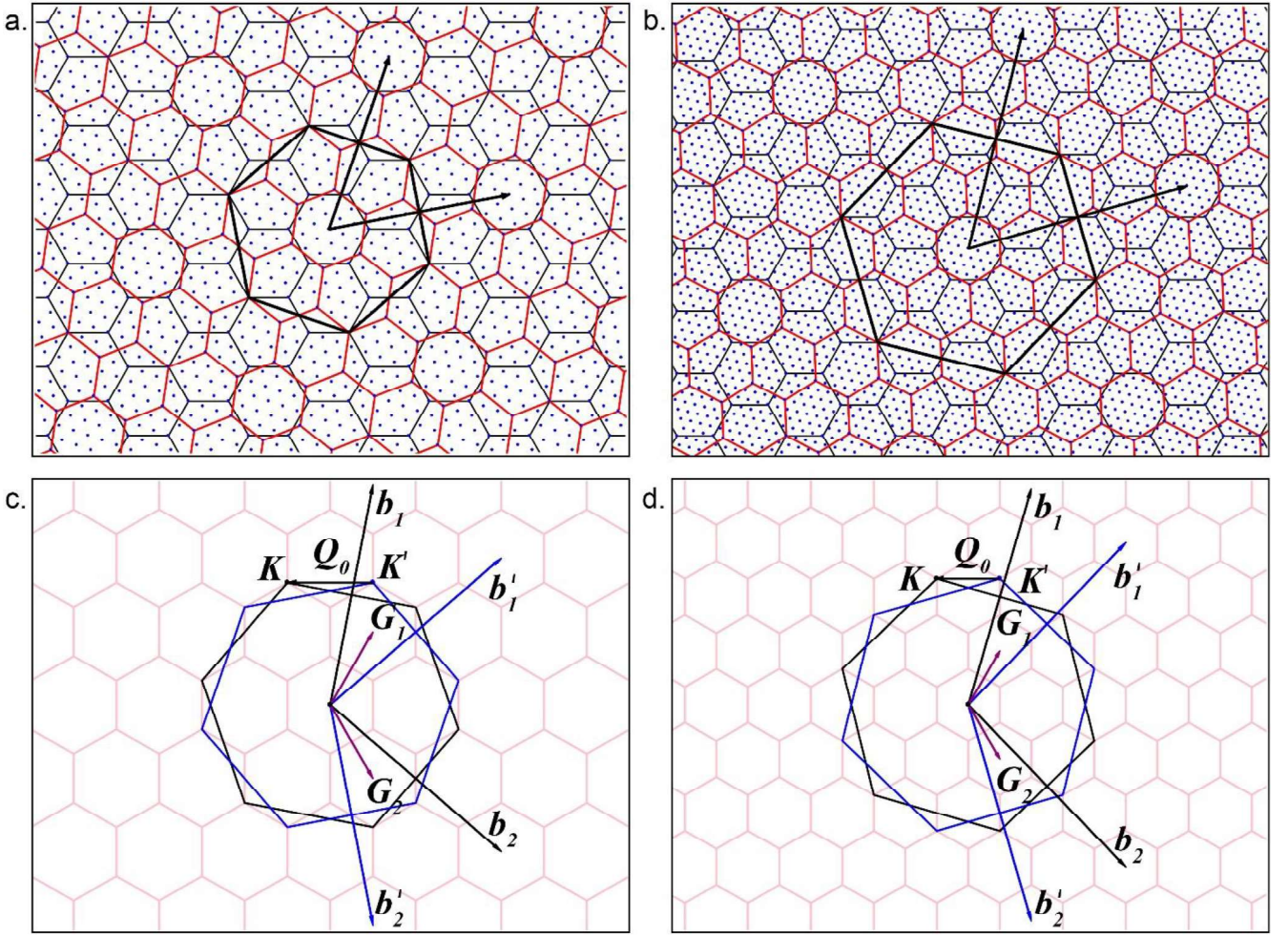

FIG. 1. The first two commensurate bilayer structures (2,1) and (3,1) in the direct (a, b) and reciprocal (c, d) spaces. (a-b) Large hexagons and arrows show the unit cell and the basis translations of superlattices, respectively. Dots represent the nodes of the primitive hexagonal lattice; the positions of carbon atoms in both the top and bottom layers coincide with some of the nodes. The existence of this parent lattice simplifies the calculation of the matrix elements of the bilayer Hamiltonian (5). The translation length of the parent lattice is $\sqrt{T}$ times shorter than the radius of the carbon hexagon. (c-d) Superposition of the top and bottom layer FBZs on a honeycomb hexagonal lattice reflecting the translational symmetry of the folded reciprocal space.

The choice of basis functions in the form (4) allows us to describe the bilayer Hamiltonian $H^b$ in a general form. Its matrix elements read:

$$H^b_{\zeta,i;\xi,j} = \frac{1}{N} \sum_{R_\zeta, R_\xi} e^{I[(q-Q_i)\cdot R_\zeta - (q-Q_j)\cdot R_\xi]} u(R_\zeta, R_\xi), \quad (5)$$

where $R_\zeta \neq R_\xi$ and $u(R_\zeta, R_\xi)$ is the hopping integral. We define the function $u(R_\zeta, R_\xi)$ following Refs. [15,30]. If both sites $R_\zeta$ and $R_\xi$ belong to the same layer and are the nearest neighbors, then $u(R_\zeta, R_\xi) = \gamma$, where $\gamma = 3$ eV is the graphene hopping coefficient [1,29], otherwise, $u(R_\zeta, R_\xi) = 0$. To calculate $u(R_\zeta, R_\xi)$, when atoms with the coordinates $R_\zeta$ and $R_\xi$ belong to the different layers one can use an approximate equation

$$u(\rho, d) = \gamma_c \exp\left[-(\sqrt{\rho^2 + d^2} - d)/\lambda\right],$$

where $\rho$ is the in-plane distance between these atoms, $d = 0.34$ nm is the inter-layer distance, $\lambda = 0.045$ nm is a characteristic wave length, $\gamma_c = 0.48$ eV is the interlayer hopping coefficient [15,31].

Note that it is convenient to index the rows and columns of the Hamiltonian matrix $H^b$ in such a way that results in $2T$ shifted blocks (1) on the matrix diagonal. Each of these blocks corresponds either to the upper or to the lower layer (they have different basis translations) and is obtained from Eq. (1) by the substitution $q \to q - Q_j$. When there is no inter-layer coupling, such a quasi-diagonal Hamiltonian describes a purely formal procedure of zone folding, as the result of which four bands of bilayer graphene are folded into $4T$ bands in common FBZ that is $T$ times smaller. The calculation of the sum (5) in the commensurate case is also simplified by the fact that the positions of both the top and bottom layers are projected onto the same parent primitive hexagonal lattice, the period of which is $\sqrt{T}$ times smaller than the distance between positions $A$ and $B$ in graphene. Therefore, the function $u(R_\zeta, R_\xi)$ takes on a discrete set of values. The number of different values in this set is equa to the number of coordination circles that surround a node of the parent lattice within the distance $\rho < \rho_0$, where $\rho_0$ is the chosen cut-off.

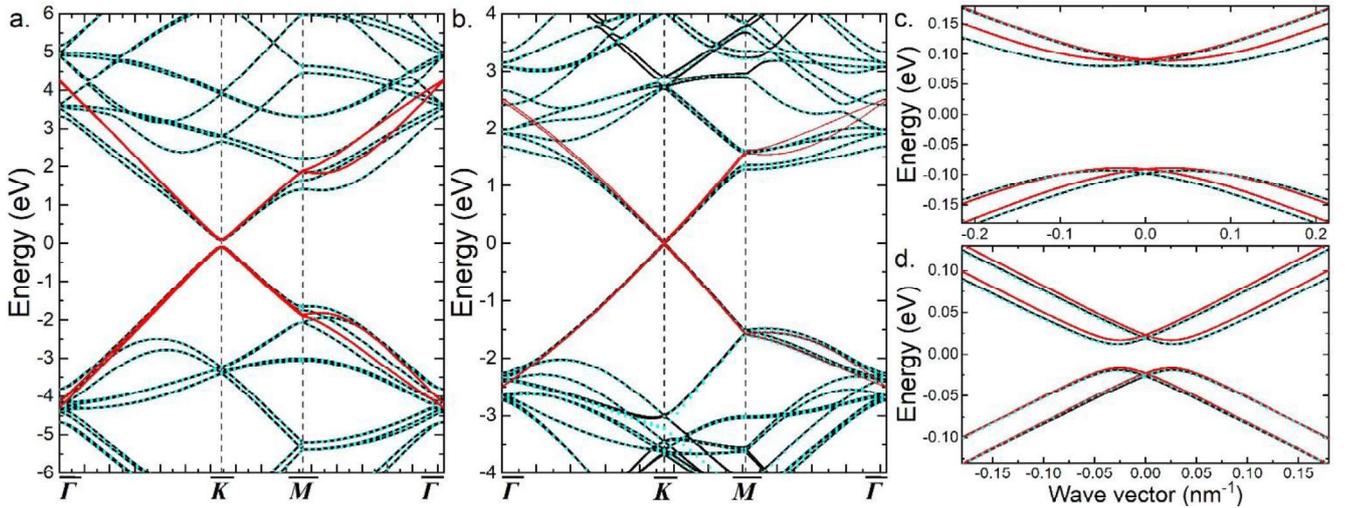

FIG. 2. Band structures calculated using the complete and effective Hamiltonians. The electron spectra are calculated within a new reduced FBZ. The point $\bar{\Gamma}$ is at its center, the point $\bar{K}$ is one of the vertices and coincides with the point $K$, the point $\bar{M}$ lies in the middle of the hexagon edge. The dispersion curves obtained within the Hamiltonian (5) are shown in black, and the ones calculated within the effective Hamiltonian (6) are shown in red. The light blue dashed line shows the spectrum of the effective Hamiltonian $H_{24}$. (a-b) The band structures for TBLG (2,1) and (3,1) are shown in the ranges $\pm 6$ and $\pm 4$ eV, respectively. On the scale of (b) there is no visible difference in the range from -3 eV to 2.5 eV between the spectra calculated within the complete Hamiltonian and the effective one $H_{24}$. (c-d) The band structure calculated near the $K$ point for the same superstructures. A gap between the bands typical of commensurate TBLG appears [15,32]. On the scale of (c,d) the spectrum of $H_{24}$ coincides with the one of the complete Hamiltonian. The origin in (c,d) is chosen in the $\bar{K}$ point. To the left from the origin, the curves in the direction $\bar{\Gamma} - \bar{K}$ are shown, while to the right from the origin, the calculations in the direction $\bar{K} - \bar{M}$ are presented.

Note that, calculating the elements of the matrix (5) the rows of which are numbered by a pair of indices $\zeta, i$ and the columns are numbered by a pair of indices $\xi, j$, one can and should limit the calculations to a unit cell of the commensurate TBLG. In the emerging bilayer superstructure, each sublattice additionally splits into $T$ sub-sublattices, whose atoms are equivalent with respect to superstructure translations. Since for atoms belonging to the same sub-sublattice with the coordinates $\mathbf{R}_j$ scalar products $\mathbf{Q}_i \mathbf{R}_j$ do not depend on the index $j$, when calculating the matrix elements (5) it is sufficient to consider only one site with indices $\zeta$ or $\xi$ (does not matter which). The range of values of the second index is limited by the fact that for sufficiently distant atoms, the function $u(\mathbf{R}_\zeta, \mathbf{R}_\xi)$ equals zero. Naturally, when using this approach, the number $N$ in Eq. (5) should be replaced with $T$.

Figure 2 (panels a and b) shows the band structures of commensurate patterns (2,1) and (3,1), in which the layers are rotated relative to each other around a common 6-fold axis. Similarly, other less symmetrical constructions can be considered, for example, the case when the relative rotation occurs around a common 3-fold axis, passing in one layer through the center of a hexagon, and in the other one through a vertex. This type of layers packing is called $AB$ or Bernal stacking [2,15,32]. For a given twist angle $\theta$, the bilayer structures corresponding to various stackings differ from each other only by a small shift of one of the layers. The latter can always be expressed as a basis translation of the parent lattice [see Fig. 1 (a-b)].

Note that regardless of the chosen type of stacking, the scheme of zone folding in the reciprocal space does not change at all, and the corresponding spectra turn out to be very close [15]. With a decrease of the twist angle $\theta$, the difference between the spectra of different commensurate structures with the same $\theta$ vanishes [15,21], as the arrangement of the atoms of one layer relative to the atoms of the other layer becomes uniformly random, just as it happens in the incommensurate case.

For calculation of the complete band structure, one can also use an equivalent Hamiltonian (see Appendix B) constructed in a basis of other $4T$ wave functions defined on $4T$ sub-sublattices. This Hamiltonian has the same dimensions, and calculating its matrix elements is a bit simpler. Nevertheless, the proposed SBWF formalism strongly simplifies the consideration of the low-energy part of the spectrum. As we show, this part of the spectrum can be obtained within the framework of simple effective Hamiltonians proposed below.

Note that due to increased translational symmetry of the reciprocal space, the vertices of the first Brillouin zones rotated relative to each other become equivalent (see Fig. 1, panels c and d). The latter leads to a strong coupling between the lowest-energy bands of the layers at the corresponding points. Let us denote the vector translating $K'$ into $K$ as $\mathbf{Q}_0$. Then, in order to describe the low-energy electron spectrum of the considered commensurate bilayer, it is necessary to take into account two pairs of SBWFs (4). The first pair with $\mathbf{Q} = 0$ is for the top layer and the second pair with $\mathbf{Q} = \mathbf{Q}_0$ is for the bottom one.

The nodes of the commensurate moiré pattern form a primitive periodic hexagonal lattice in reciprocal space, and the vector $\mathbf{Q}_0$ can be expressed as an integer linear combination of this lattice basis translations $\mathbf{G}_1$ and $\mathbf{G}_2$. For a commensurate superstructure $(h, k)$, these translations read as follows:

$$\mathbf{G}_1 = \frac{\Delta \mathbf{b}_1 + 2\Delta \mathbf{b}_2}{3k}, \quad \mathbf{G}_2 = -\frac{2\Delta \mathbf{b}_1 + \Delta \mathbf{b}_2}{3k},$$

where $b_1, b_2$ ($b_1', b_2'$) are the basis vectors of the top (bottom) layer reciprocal lattice and $\Delta b_1 = b_1 - b_1'$, $\Delta b_2 = b_2 - b_2'$. Accordingly, the vector $Q_0$ translating $K'$ into $K$ is found as $-k(G_1 + G_2)$.

Note that the vector $Q_0$ can be also expressed as an integer linear combination of the vectors $b_1, b_2$ and $b_1', b_2'$. A shift by a vector of the top (bottom) layer reciprocal space only changes the phase of the Bloch function. Therefore, when calculating the band structure one can neglect the dependence of $Q_0$ on $b_1'$ and $b_2'$, let us denote such a vector as $Q_s$. Thus, we obtain the first effective Hamiltonian

$$H^{eff} = \begin{pmatrix} H(q) & V(q, Q_s) \\ V^*(q, Q_s) & H'(q - Q_s) \end{pmatrix}, \quad (6)$$

where the non-diagonal blocks are calculated as

$$V_{\alpha,\alpha'} = \frac{1}{T} \sum_{R_\alpha, R_{\alpha'}} e^{i[q \cdot R_\alpha - (q - Q_s) \cdot R_{\alpha'}]} u(R_\alpha, R_{\alpha'}), \quad (7)$$

$H(q)$ and $H'(q)$ are the Hamiltonians with the form (1) constructed for still and rotated layers respectively, $\alpha = A, B$ and $\alpha' = A', B'$.

In the case of the pattern (2,1) (its reciprocal space is shown in Figure 1c), the vector translating $K'$ into $K$ is $-(b_1 + b_2')$. Therefore, as the vector $Q_s$ one can take the vector $-b_1$. For the pattern (3,1) (see Fig. 2c), this vector is $2(b_1 + b_2') - (b_1' + b_2)$, then $Q_s = 2b_1 - b_2$. The spectra of both structures, calculated within the Hamiltonian (6), are shown in Fig 2 by red lines.

Note that the spectrum of the complete Hamiltonian for any commensurate superstructure is periodically repeated in the reciprocal space spanned by the basis vectors $G_1$ and $G_2$. The constructed effective Hamiltonian (6) does not have such a translational symmetry and approximates the complete one only in the vicinity of the point $K$ coinciding with the vertex $\bar{K}$ of the cell of the folded reciprocal space. The accuracy of such an approximation of the low-energy spectrum decreases when approaching the points $\bar{\Gamma}, \bar{M}$ (see Fig. 2) and $-\bar{K}$.

Let us now consider how a more complex effective Hamiltonian can be constructed, which approximates well the low-energy spectrum within a cell of the folded reciprocal space for any commensurate bilayer structure. Recall that the previous effective Hamiltonian (6) is constructed in such a way that when there is no inter-layer coupling, the peaks of the considered pair of Dirac cones exactly coincide with the point $\bar{K}$ of the reduced FBZ. The next effective Hamiltonians can be constructed using the same idea. Namely, in order to approximate the spectrum within a unit cell with six vertices, we choose shift vectors in SBWFs in such a way that a pair of Dirac cones [which correspond to the blocks $H(q - Q_i)$ and $H'(q - Q_j')$, respectively] appears in every vertex of the cell. The basis of the Hamiltonian proposed, which we denote as $H_{24}$, consists of 24 shifted Bloch functions. Figure 2 (a,b) shows the spectra of the complete Hamiltonian (black solid lines) and the spectra of the effective Hamiltonian $H_{24}$ (light blue dashed lines) calculated for the structures (2,1) and (3,1). In the low-energy region, the difference between the dispersion curves is not noticeable on the scale of Fig. 2. For these structures and several following ones (which we analyzed numerically), the effective Hamiltonian $H_{24}$ reproduces the low-energy spectra of the complete Hamiltonian with an accuracy of up to 10 meV.

Note that the constructed Hamiltonian $H_{24}$ consists of two weakly coupled blocks of two times smaller dimension. The basis functions of the first block are chosen in such a way that three pairs of Dirac cones originating from the top and bottom layers appear in three translationally equivalent vertices of the hexagonal cell. From a practical point of view, it is convenient to choose the reduced cell that contains the points $K$ and $K'$ [see Fig. 1 (c-d)]. Then the basis of the first block includes the basis functions of the Hamiltonian (6), as well as 8 functions obtained from these four by using two minimal basis translations that shift the points $K$ and $K'$ to the vertices of the same reduced cell. For example, for the structure (3,1) this basis consists of SBWFs with translations $Q_i = [0, G_2, G_1 + G_2]$ for the top layer, and SBWFs with translations $Q_j' = [0, -G_1, -G_1 - G_2]$ for the bottom one.

To clarify the origin of the 12 functions of the second block, we note that the three remaining vertices of the considered cell are not translationally equivalent to the first three ones. However, each shifted Hamiltonian (1) produces Dirac cones that simultaneously belong to both sublattices of the honeycomb lattice of the folded reciprocal space. Therefore, the basis functions are shifted only by integer linear combinations of the vectors $G_1$ and $G_2$.

To determine the shift vectors for the functions of the second block, note that Hamiltonian (6) also approximates the spectrum of the full Hamiltonian in the vicinity of the point $-K$, which belongs to another cell of the folded zone, obtained from the considered one by inversion of the reciprocal space. These two cells are translated into each other in the folded reciprocal space by a vector $\tilde{G}$. For the structures shown in Fig. 1c and 1d, this vector is expressed as $\tilde{G} = 2(G_2 - G_1)$. Accordingly, the functions of the second block can be obtained from the first 12 functions simply translating them by $\tilde{G}$. The fact that the length of the minimal basis translation of the folded reciprocal space $G$ is substantially less than the length of $\tilde{G}$ explains the weak coupling between the 12x12 blocks of the Hamiltonian $H_{24}$. The latter fact also explains the weak repulsion of low-energy bands originating from the different blocks (see the spectra near the points $\bar{\Gamma}$ and $\bar{M}$ in Fig. 2a-b). The ratio $G/\tilde{G}$ decreases with the twist angle. As a result, the coupling between the Dirac cones located in the translationally non-equivalent vertices of the cell also weakens.

Within the framework of the proposed theory, it is easy to clarify the flattening of the lowest-energy bands in TBLG with a small twist angle. Even comparing Figures 2a and 2b, one can see that with a decrease in $\theta$, the distance between the bands (the maximum one is near $\bar{\Gamma}$) has decreased. Figure 3 shows the low-energy regions of spectra calculated within the Hamiltonian $H_{24}$ for structures (15,1) and (27,1), where one can see the same tendency. Nevertheless, the band structure produced by the Hamiltonian $H_{24}$ is too complex for a qualitative consideration of the occurring phenomenon. Therefore, in Fig. 3a we schematically show the rearrangement in the spectrum resulting from the coupling between only two translationally equivalent pairs of Dirac cones. These cones correspond to a Hamiltonian with the basis of 8 SBWFs that have the following shift vectors: $Q_i = [0, Q_0]$ and $Q_j' = [0, -Q_0]$. As shown in the figure, due to Van der Waals interaction, the dispersion curves merge and repel. As a result, the intersections between the cones split into several bands and the lowest-energy ones are pushed by the bands above towards the Fermi level $E = 0$.

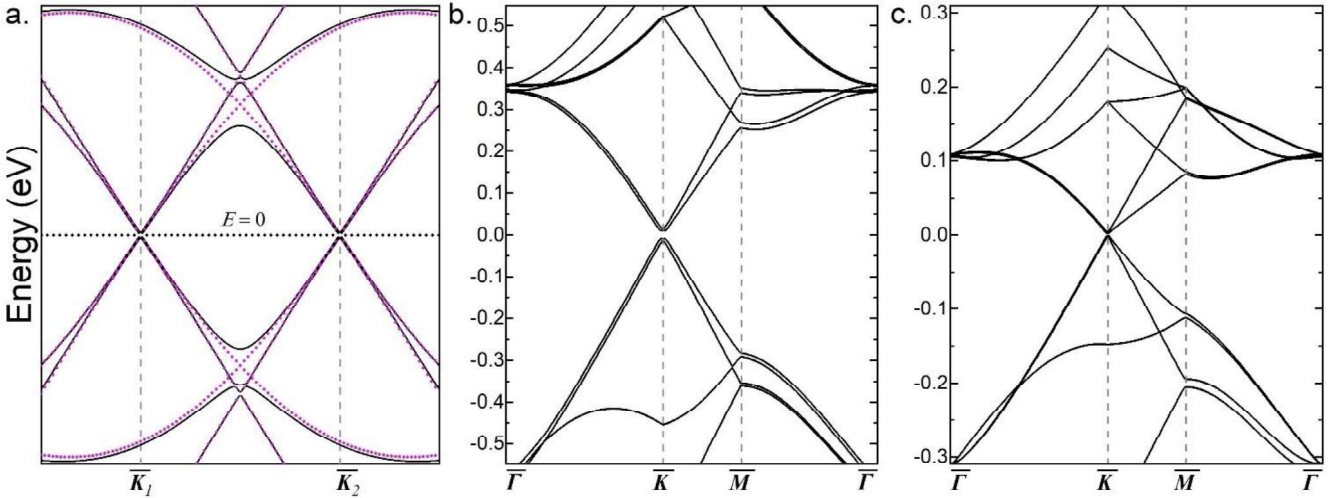

FIG. 3. Flattening of the lowest-energy bands in TBLG with a small twist angle within the framework of zone folding theory. (a) Schematic band structure produced by the coupling between two pairs of Dirac cones. The cone location points $\overline{K}_1$ and $\overline{K}_2$ are separated by a minimum translation with length $G$. Purple dotted lines show the band structure when the coupling is switched off. The cones of the top and bottom layers are rotated by an angle $\theta$ and, therefore, coincide only in the vicinity of the points $\overline{K}_1$ and $\overline{K}_2$. The band structure (calculated with the account of Van der Waals interaction between the layers) is shown by black solid lines. This interaction leads to the merging and repulsion of previously intersecting dispersion curves. (b-c) Spectra of the effective Hamiltonian $H_{24}$ calculated for commensurate structures (15,1) and (27,1). The corresponding twist angles θ are 6.40° and 3.61°. The maximum distance between the lowest energy bands (which is near the point $\overline{\Gamma}$) decreases from 1 to 0.4 eV.

Note that the maximum distance between the bands under consideration can be estimated as $GV_F$, where $V_F = \sqrt{3}\gamma a/2$ is the Fermi velocity in the graphene monolayer and $a$ is the graphene lattice constant. The length $G$ tends to 0 together with the twist angle $\theta$, and this fact clearly explains the band flattening and the corresponding decrease in the gap between the bands.

Concluding the discussion of this problem, we note that although the flattening mechanism within the proposed approach becomes practically obvious, a more detailed analysis of the EDoS, including the consideration of "magic angles", is clearly beyond the scope of this work.

We also note that the constructed effective Hamiltonian $H_{24}$ is only suitable to describe the lowest-energy region of the spectrum. To describe the spectrum in a region with a width of $\pm \Delta E$, the basis of such a Hamiltonian should include additional shifted Bloch functions that correspond to Dirac cones pairs falling onto the area with a radius $Q_r$ and center in $\overline{\Gamma}$, where $Q_r \leq \Delta E/V_F$. When the interlayer coupling is switched off, in the folded reciprocal space, these functions exactly correspond to the bands whose energies lie within the range of $\pm \Delta E$.

Below, within the proposed formalism of shifted Bloch functions, we will consider the construction of simpler effective Hamiltonians for the incommensurate TBLG and incommensurate DWCNTs.

## III. INCOMMENSURATE BILAYER STRUCTURES

As mentioned before, a bilayer graphene with an arbitrary twist angle $\theta$ is an aperiodic structure. However, the vectors $\Delta \boldsymbol{b}_1$ and $\Delta \boldsymbol{b}_2$ always correspond to a periodic hexagonal lattice in the direct space. Using the orthogonality relations $\boldsymbol{C}_i^M \cdot \Delta \boldsymbol{b}_j = 2\pi \delta_{ij}$, where $\delta_{ij}$ is the Kronecker delta,

one can obtain the basis translations of this so-called moiré lattice:

$$\boldsymbol{C}_1^M = \frac{\boldsymbol{a}_1' - \boldsymbol{a}_1}{4\sin^2\frac{\theta}{2}}, \boldsymbol{C}_2^M = \frac{\boldsymbol{a}_2' - \boldsymbol{a}_2}{4\sin^2\frac{\theta}{2}},$$

where $\boldsymbol{a}_1, \boldsymbol{a}_2$ and $\boldsymbol{a}_1', \boldsymbol{a}_2'$ are the basis graphene translations in the top and bottom monolayers, respectively. The period of the moiré lattice is

$$C^M = |\boldsymbol{C}_1^M| = |\boldsymbol{C}_2^M| = \frac{\sqrt{3}a_0}{2\sin\left(\frac{\theta}{2}\right)},$$

where $a_0 = 0.142$ nm is the carbon bond length, and the area of the moiré unit cell is found as

$$S^M = |\boldsymbol{C}_1^M \times \boldsymbol{C}_2^M| = \frac{3\sqrt{3}a_0^2}{8\sin^2\left(\frac{\theta}{2}\right)}.$$

In the vicinity of moiré lattice nodes, the TBLG structure is approximately the same; however, due to incommensurability between the lattice and graphene monolayers, a slight shift appears when translating from one lattice node to the other (see Fig. 4a).

It is clear that the electron spectrum of incommensurate TBLG can be approximated by the spectrum of the above-considered commensurate TBLG, since the only parameter characterizing TBLG structure (which is the twist angle) can change continuously. Nevertheless, below we consider a simple effective Hamiltonian constructed specifically for incommensurate TBLG. The advantage of such a Hamiltonian is that it can be easily adapted for the case of incommensurate DWCNT. The structures of commensurate and incommensurate double-walled tubes are qualitatively different. It is simply impossible to construct such a commensurate tube approximating an incommensurate one, since the DWCNT structure is simultaneously characterized by four parameters: two diameters and two chiral angles.

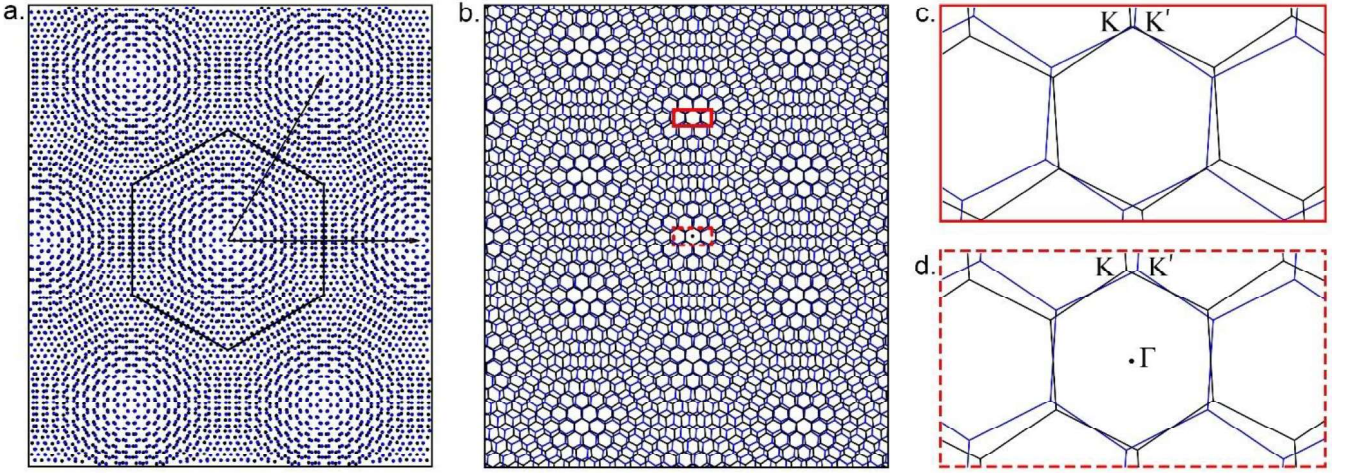

FIG. 4. The direct and reciprocal spaces of the incommensurate TBLG. (a) Two superimposed hexagonal lattices with a twist angle $\theta = 3.5°$. The arrows show the basis translations of the moiré lattice. Near the nodes, the TBLG structure is practically the same; however, the translations of the moiré lattice connect structurally nonequivalent points of TBLG (b) Superimposed reciprocal lattices of the top and bottom monolayers; $\theta = 3.5°$. (c-d) Selected regions of the panel (b) at a larger scale. Far from the origin, points of $K$ and $K'$ types can be arbitrarily close; an example is shown in panel (c). (d) Around the point $\Gamma$ there are three pairs of $K$ and $K'$ points.

Now, let us consider the differences between reciprocal lattices of an incommensurate bilayer and a commensurate one. In the infinite reciprocal space of the incommensurate TBLG, $K$ and $K'$ points can be arbitrarily close, corresponding, at first glance, to a strong inter-layer coupling (see Fig. 4b). However, with the exception of only 3 pairs of $K$ and $K'$ points, the coupling between the corresponding modes is quite weak, since the matrix elements (7) decay exponentially with the increasing distance $|q - Q_s|$ in the reciprocal space (see Appendix C). In other words, if we superimpose two incommensurate reciprocal spaces (of TBLG or DWCNT), then near the points, where vertices $K$ and $K'$ are close to each other and locate far from the origin the corresponding inter-layer interactions can be neglected [33].

Let us discuss how to construct in the incommensurate case the effective Hamiltonian valid in the vicinity of the point $K_1 = 1/3(b_1 - b_2)$. Note that there are two other translationally equivalent nodes of the graphene FBZ which are close to the point $\Gamma$: $K_2 = K_1 + b_2$ and $K_3 = K_1 - b_1$. Near each of these points, there are vertices $K'_1$, $K'_2$ and $K'_3$, respectively. If only pairwise interactions between the modes with the wave vectors $K_i$ and $K'_i$, are taken into account, then the effective Hamiltonian can be constructed using 4 pairs of SBWFs. The first and the second pairs (for the top and bottom layers, respectively) are both with $Q_s^0 = 0$. The remaining two pairs of Bloch functions are also for the lower layer and have the following shift vectors $Q_s^1 = -b_2$ and $Q_s^2 = b_1$. Accordingly, the effective Hamiltonian for the incommensurate TBLG reads

$$\begin{pmatrix} H(q) & V(q,0) & V(q,b_1) & V(q,-b_2) \\ V^{*T}(q,0) & H'(q) & 0 & 0 \\ V^{*T}(q,b_1) & 0 & H'(q-b_1) & 0 \\ V^{*T}(q,-b_2) & 0 & 0 & H'(q+b_2) \end{pmatrix}, \quad (8)$$

where $H(q)$ and $H'(q)$ are matrices obtained from Eq. (1) by substituting the corresponding basis vectors of the direct lattice, the superscript $T$ denotes the operation of matrix transposition.

Note that Hamiltonian (8) can be rewritten using the reciprocal space vectors $\Delta b_1$ and $\Delta b_2$ of a moiré lattice, which is in many respects similar to the lattice of the folded reciprocal space arising in the commensurate case. Then the Hamiltonian (8) turns out to be very similar to one of the weakly coupled blocks of the Hamiltonian $H_{24}$ described at the end of the previous section. The difference is due to the fact that the vector connecting the points $K$ and $K'$ is not a translation of the moiré lattice, and the reduced dimension of the matrix (8) is explained by the asymmetry of this Hamiltonian with respect to the permutation of the upper and lower layers.

It may also seem that Eq. (7) can still be used to calculate the matrix elements $V(q, Q_s^j)$ of the Hamiltonian (8), where it is sufficient to replace the triangulation factor $T$ with the number of hexagons that fall on a moiré unit cell with an area $S^M$. However, the translations $C_1^M$ or $C_2^M$ along the moiré lattice connect non-equivalent points of incommensurate TBLG, since the latter is incommensurate with both graphene layers. As a result, an additional small phase at the points near the moiré lattice nodes can significantly impact the value of the matrix elements and this phase vanishes only when averaged over an *infinite* incommensurate bilayer lattice.

As our numerical analysis shows, to evaluate $V(q, Q_s)$ one should perform the calculations at least over several tens of moiré pattern cells, which dramatically affects the computational efficiency. The situation is a bit simplified by the fact that due to the incommensurability of the layers, the elements $V_{\alpha\alpha'}(q, 0)$ practically do not depend on the sublattice indices $\alpha$ and $\alpha'$, and moreover, they are practically real. Thus, $V_{\alpha\alpha'}(q, 0) \approx v(q)$, where the function $v(q)$ can be calculated using the simplified equation

$$v(q) = \frac{1}{N} \sum_{R_\alpha, R_{\alpha'}} \cos[q \cdot (R_\alpha - R_{\alpha'})] u(R_\alpha, R_{\alpha'}), \quad (9)$$

in which the summation is performed over any two sublattices belonging to different TBLG layers. See more detail on deducing Eq. (9) in Ref. [27]. Note that the matrix elements $V(q, Q_s)$ have the following property:

$$V_\alpha(q, Q_s) = \exp(I Q_s R_\alpha) v(q - Q_s), \quad (10)$$

where the vector $Q_s$ can be any translation of the top layer reciprocal lattice, in particular, $b_1$ or $-b_2$. As $R_\alpha$ one can choose the coordinates of any of the atoms belonging to the sublattice $\alpha$, since it does not change the phase in Eq (9). Thus, to obtain all the non-diagonal elements of the Hamiltonian (8) one should calculate the function $v(q)$ and then determine the complex factors according to Eq. (10). In the explicit form, the non-diagonal blocks of Hamiltonian (8) are found as $V(q, b_1) = Cv(q - b_1)$ and $V(q, -b_2) = C^* v(q + b_2)$, where $C = \begin{pmatrix} \varepsilon & \varepsilon \\ \varepsilon^* & \varepsilon^* \end{pmatrix}$ and $\varepsilon = \exp(I 2\pi/3)$.

The tight binding theory developed above for bilayer graphene can be generalized to the case of DWCNTs. In order to do so, one should express the graphene primitive translations $a_1$ and $a_2$ in cylindrical coordinates, where the first and second components of a vector correspond to the projections onto the circumference and the longitudinal axis of the tube, respectively (see Appendix A). Following Ref. [27], we define the first component of the cylindrical translation as a fraction of the SWCNT perimeter. With this definition, the first component $\mu$ of the wave vector $q = (\mu, k)$ becomes a dimensionless integer; it numbers the cutting lines in the unrolled reciprocal space of SWCNTs, while the second component $k$ is the projection of the wave vector along the cutting lines [17].

Let us note that all the structurally identified DWCNTs so far are incommensurate [34]; thus, we consider only this case below. Unlike completely equivalent graphene layers, the inner and outer tubes of DWCNTs are not equivalent to each other. Therefore, the lowest-energy dispersion curves of the outer and inner tubes can be considered within the framework of the Hamiltonian in a basis of 12 SBWFs (two for each pair of $K$ and $K'$ points). An alternative and simpler option is to use two different Hamiltonians for the outer and inner tubes. Then each of them turns out to be similar to the Hamiltonian of incommensurate bilayer graphene considered above. For example, the effective Hamiltonian for the inner tube reads as follows:

$$\begin{pmatrix} H_{in}(q) & Ev(q) & Cv(q-b_1) & C^*v(q+b_2) \\ Ev(q) & H_{out}(q) & 0 & 0 \\ C^{*T}v(q-b_1) & 0 & H_{out}(q-b_1) & 0 \\ C^T v(q+b_2) & 0 & 0 & H_{out}(q+b_2) \end{pmatrix}, \quad (11)$$

where $E = \begin{pmatrix} 1 & 1 \\ 1 & 1 \end{pmatrix}$, the blocks $H_{in}(q)$ and $H_{out}(q)$ are obtained from the Hamiltonian (1) by the described above substitution of the basis vectors of the direct and reciprocal lattices. The eigen energies of the blocks $H_{in}(q)$ and $H_{out}(q)$ are respectively equal to $E_{in}^\pm = \pm|f_{in}(q)|$ and $E_{out}^\pm = \pm|f_{out}(q)|$. The vectors $b_1$ and $b_2$ are the basis vectors of the inner tube reciprocal space (see Appendix A).

The matrix elements $v(q)$ in the Hamiltonian (11) have the same properties as the elements of the incommensurate TBLG [27]. They are also real and practically do not depend on the indices $\alpha$ and $\alpha'$, with the only difference that for the case of DWCNT in Eq. (9) the number $N$ should be replaced with $\sqrt{N_{in} N_{out}}$, where $N_{in}$ and $N_{out}$ are the numbers of hexagons in the inner and outer tubes, respectively. To preserve the accuracy when calculating $v(q)$ according to (9), it is necessary to consider a DWCNT with a length exceeding several tens of moiré periods along the tube axis.

Let us denote $E^+(E^-)$ as the energy of the DWCNT band originating from the inner SWCNT conduction (valence) band. Expanding the proposed effective Hamiltonian (11) into the series up to the second order of smallness in $v(q)$, one can obtain the energy shifts for the conduction band $\Delta E_{in}^+(q) \equiv E^+ - E_{in}^+$ and the valence band $\Delta E_{in}^-(q) \equiv E^- - E_{in}^-$ of the inner tube:

$$\Delta E_{in}^+(q) = \sum_{j=1}^{3} \frac{v_j^2 (1+\cos\varphi_{in})(1+\cos\varphi_{out,j})}{|f_{in}| - |f_{out}|_j} + \frac{v_j^2 (1+\cos\varphi_{in})(1-\cos\varphi_{out,j})}{|f_{in}| + |f_{out}|_j}, \quad (12)$$

$$\Delta E_{in}^-(q) = -\sum_{j=1}^{3} \frac{v_j^2 (1-\cos\varphi_{in})(1-\cos\varphi_{out,j})}{|f_{in}| - |f_{out}|_j} + \frac{v_j^2 (1-\cos\varphi_{in})(1+\cos\varphi_{out,j})}{|f_{in}| + |f_{out}|_j}. \quad (13)$$

Eqs. (12-13) yield the energy shift for the direct electronic transition in the inner nanotube:

$$\Delta E_{in}(q) = \sum_{j=1}^{3} \frac{2v_j^2 (1+\cos\varphi_{in} \cos\varphi_{out,j})}{|f_{in}| - |f_{out}|_j} + \frac{2v_j^2 (1-\cos\varphi_{in} \cos\varphi_{out,j})}{|f_{in}| + |f_{out}|_j}. \quad (14)$$

In Eqs. (12-14) all the quantities depend on the wave vector: $v_j = \{v(q), v(q-b_1), v(q+b_2)\}$; $\varphi_{in}$ and $\varphi_{out,j}$ are the phase shifts between the sublattices of the inner and outer tubes, respectively (see Eq. 2); the values of $|f_{out}|_j$ and $\varphi_{out,j}$ are calculated with the same wave vector shifts $\{0, b_1, -b_2\}$. Obviously, the obtained expansions are applicable provided that $v \ll ||f_{in}| - |f_{out}||$.

The effective Hamiltonian for the outer tube is obtained from Eq. (11) by a simple permutation of the indices *in* and *out*. It is also necessary to replace the vectors $b_i$ with the vectors $b_i'$. The energy shifts for the bands of the outer tube are obtained from Eqs. (12-14) by the same substitutions. Note that Eqs. (12-14) can also be obtained within the framework of the perturbation theory developed in Ref. [27].

In the next section, the developed theory of the band structure in incommensurate bilayer systems will be applied to calculate the energies of optical transitions in DWCNTs and, in particular, to calculate the energies of novel inter-tube transitions. This type of optical transitions has been discovered recently and theoretically considered only for a few DWCNTs [26–28].

## IV. OPTICAL TRANSITIONS IN DOUBLE-WALLED CARBON NANOTUBES

The theory developed above and Eq. (14), in particular, can be used to calculate the energy shifts of optical transitions in DWCNTs. In this case, one should take into account that due to many-electron interactions, the energy of optical transitions in SWCNTs increases by several hundred meV compared to the predicted ones by the tight-binding approach [35] and the conduction band is replaced by an exciton band [36]. However, the formation of a DWCNT from two pristine SWCNTs practically does not affect these effects, since the binding energy of excitons significantly exceeds the energy of the Van der Waals interaction between the SWCNTs. For the first time, perturbation theory combined with TBA was successfully applied in Ref. [33]. Using this approach, the authors calculated the energy shifts due to inter-layer coupling and compared them to the experimentally observed shifts in the positions of absorption peaks. Our theory can also be applied to describe the energy shift of electron transition when a DWCNT is formed from SWCNTs. The optical transition energy in SWCNTs cannot be calculated using NN TBA, and one should use experimental data or a semi-empirical equation from Ref. [37] approximating the data on ~200 SWCNTs with the diameters ranging from 1.3 nm to 4.7 nm. Note that in some relatively rare cases, the error of the formula (1) from Ref. [37] can reach up to 40-50 meV. When analyzing the experimental data, one should also consider the screening effect in DWCNTs [33,38–40], which leads to a redshift in the optical transition energies by 50-60 meV.

Let us recall that the electron transitions observed in the optical spectra of SWCNTs originate from points in reciprocal space called van Hove singularities. Near the $K$ points, the VHS coordinates can be found as [27]
$$\boldsymbol{q}_p = \boldsymbol{K} + \boldsymbol{P},$$
where $\boldsymbol{K}$ can be one of three vectors $\{\boldsymbol{K}_1 = 1/3(\boldsymbol{b}_1 - \boldsymbol{b}_2),$ $\boldsymbol{K}_2 = 1/3(\boldsymbol{b}_1 + 2\boldsymbol{b}_2)$ and $\boldsymbol{K}_3 = 1/3(-2\boldsymbol{b}_1 - \boldsymbol{b}_2)\}$, $\boldsymbol{P} = (p/3, \delta k)$, $p$ is an integer, and $\delta k$ describes a small band extremum shift along the cutting line $\mu(p)$. The higher the energy of the band is, the larger the shift $\delta k$ becomes. Positive and negative numbers $p$ which are multiples of three ($|p| = 3,6,9$) correspond to transitions in metallic SWCNTs ($M_{11}, M_{22}, M_{33}$). These transitions are split, and a positive number corresponds to a slightly higher energy of the doublet than a negative one [41]. Other positive and negative numbers ($|p| = 1,2,4,5,7,8$) index electronic transitions in semiconducting tubes ($S_{11}, S_{22}, S_{33}, S_{44}, S_{55}, S_{66}$). For the SWCNT $(n, m)$, the sign of $p$ numbers is unequivocally determined by the integer constraint for the "angular" component $\mu(p) = (n - m + p)/3$ of the vector $\boldsymbol{q}$.

As we demonstrated above, in the case of DWCNTs, it is necessary to consider the coupling between the modes of the inner and outer tubes near three $K$ points. Then, in order to find the energy shift of an optical transition with index $p$ originating from a SWCNT, one can use Eq. (14) rewritten as:

$$\Delta E(p) = \sum_{j=1}^{3} \Delta E_{in(out)}(\boldsymbol{K}_j + \boldsymbol{P}) + \Delta_s, \quad (15)$$

where $\Delta_s$ is the constant redshift occurring due to the screening effect.

In Ref. [27], we have successfully applied Eq. (15) to analyze 94 intra-tube optical transitions in 27 DWCNTs taken from Ref. [33]. With the found fitting parameters of the theory [27], the energies of these transitions have been calculated with a standard error of 18 meV, and the maximum error has been 47 meV. The only drawback of this approach has been the rather time-consuming calculation of Eq. (9).
Following the approximation [27,33], which states that atom positions in different layers can be considered approximately random relative to each other, the summation over the inner or outer tube sites in Eq. (9) can be replaced by integration. Accordingly, the interlayer matrix element is simply found as the Fourier transform of the hopping integral $u(\rho, d)$ (see Appendix C). Note that since the inter-layer distance in DWCNTs is a variable, the parametrization of the hopping integral is changed. If it is transformed (like in Refs. [27,33]) into the function $w(l) = \gamma_c \exp(-l/\lambda)$, where $l$ is the distance between the considered atoms of adjacent layers, then the integration yields a special function, which is not practical for our calculations. However, since both functions $u(\rho, d)$ and $w(l)$ are essentially phenomenological ones, then in order to obtain observable expressions, we used a Gaussian function $u(l) = \gamma_c' \exp(-\alpha l^2)$. As a result, we found the following expression to calculate the inter-layer matrix elements:

$$v(\mu, k) = \frac{\gamma_c' \pi}{S_0 \alpha} e^{-\alpha \Delta R^2} e^{-\frac{1}{4\alpha}\left(k^2 + \frac{\mu^2}{R_{in}R_{out}}\right)}, \quad (16)$$

where $R_{in}$ and $R_{out}$ are radii of the inner and outer SWCNT, $\Delta R = R_{out} - R_{in}$ is the inter-layer distance in a DWCNT, $\mu$ and $k$ are the wave vector $\boldsymbol{q}$ components, $S_0$ is the area of the graphene unit cell. The parameters $\alpha$, $\gamma_c'$ and $\Delta_s$ were determined by the method of least squares and minimization of the error in the calculations of $\Delta E$. For this fit, we used the experimental data on 94 optical transitions from Ref. [33]. Due to reasons explained in Ref. [27], the data on transitions from DWCNT (11,7)@(21,6) were excluded. As a result, we obtained the following values of the constants: $\gamma_c' = 46$ eV, $\alpha = 38$ nm$^{-2}$. The found values of the screening constant $\Delta_s$ are -55 meV and -60 meV when the energy shift $\Delta E$ is calculated for a transition originating from a metal tube and a semi-conducting one, respectively. With these parameters, the standard error in our calculations turned out to be slightly less than 18 meV, and a maximum error was 44 meV

Note that a similar expression for calculation of the inter-layer coupling (16) can be obtained by simply expanding the argument of the function $\exp(-l/\lambda) = \exp(-\sqrt{\rho^2 + \Delta R^2}/\lambda)$ up to the second order of smallness in $\rho/\Delta R$. Probably, this procedure was carried out in Ref. [33], where the related details were omitted. As a result of this expansion, the hopping integral is also expressed as a Gaussian function, but with different parameters and a different dependence on the distance $\Delta R$. Even though the standard error practically does not change within this approach, the maximum error increases up to 50 meV.

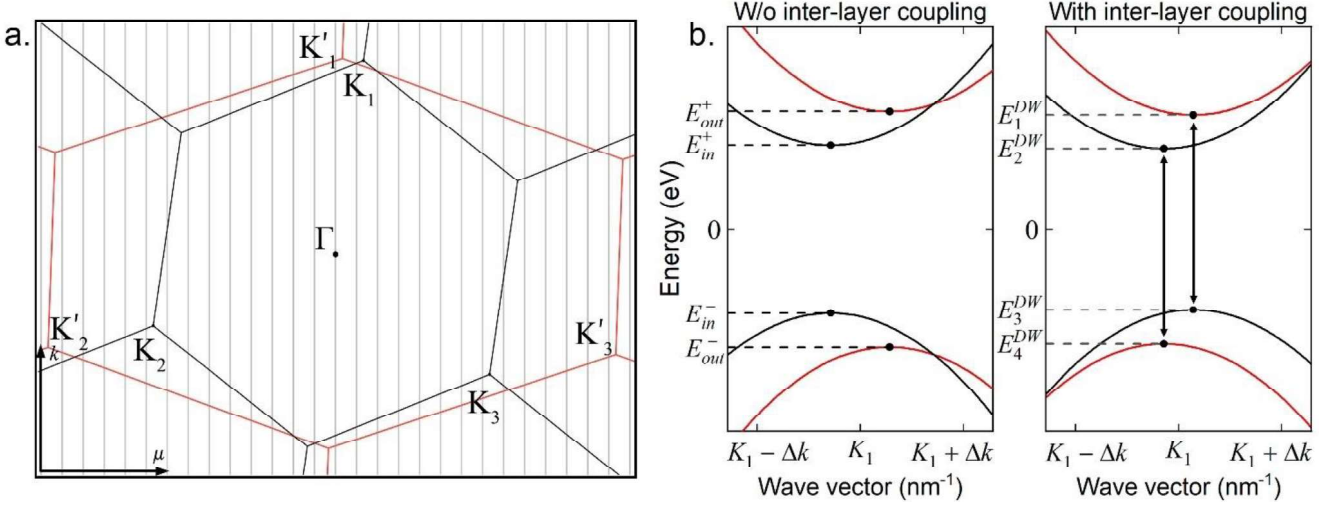

FIG. 5. Origin of inter-tube optical transitions in DWCNTs with the specific geometry. (a) Typical reciprocal space of a DWCNT, where ITTs are possible. Superimposed extended Brillouin zones of the inner and outer tubes are shown in red and black, respectively. Parallel grey lines represent the common cutting lines of the DWCNT. The projections of the points $K'_1$ and $K_1$ on these lines have very close $k$-coordinates, which corresponds to the proximity of the band extrema originating from different pristine SWCNTs. As an example, DWCNT (10,6)@(14,13) is chosen. (b) Schematic rearrangement in the DWCNT band structure caused by inter-layer coupling. Black and red solid curves correspond, respectively, to the bands of the inner and outer tubes; $\Delta k \ll K_1$, where $K_1$ is the $k$-coordinate of the point $\mathbf{K_1}$; $E^+_{in(out)}$ and $E^-_{in(out)}$ are the energies of the conduction and valence bands in non-interacting tubes; $E^{DW}_1, E^{DW}_2, E^{DW}_3, E^{DW}_4$ are the corresponding energies of DWCNT bands. Double arrows demonstrate the emerging inter-tube transitions.

Besides its accuracy and computational efficiency, the developed theory has the advantage that it allows calculating the energies of inter-tube optical transitions (ITTs), which has been theoretically predicted quite recently [27,28]. We found the conditions allowing for ITTs in Ref. [27]. We demonstrated that these transitions arise due to a rearrangement in the band structure which is only possible in DWCNTs satisfying specific geometrical criteria. More precisely, ITTs are allowed in those DWCNTs where the points $\mathbf{K_1} = (\mathbf{b_1} - \mathbf{b_2})/3$ and $\mathbf{K'_1} = (\mathbf{b'_1} - \mathbf{b'_2})/3$ are close to each other [27], i.e. these points should not be separated by more than two cutting lines in the reciprocal space, and the difference in their $k$-coordinates along the tube axis should also be small, much less than the inverse periods of the inner and outer SWCNTs. An example of a reciprocal lattice satisfying these criteria is presented in Fig 5a. When these conditions are met, the VHSs of the outer and inner tubes turn out to be close in the reciprocal space. The inter-tube coupling can modify the band structure in such a way that VHSs originating from different tubes become even closer, while the bands belonging to the same tubes move apart. This type of rearrangement is illustrated in Fig 5b. Note that the valence and conduction bands of the inner and outer tubes can be shifted differently, and the energies of a pair of ITTs do not always coincide. In the earlier study [27] we have shown that the values of dipole moment matrix elements for inter-tube and intra-tube transitions can be of the same order of magnitude. Since the intensity of a transition is proportional to the value of the dipole matrix element and EDoS [42,43], both types of electron transitions can be present in the optical spectra of DWCNTs. However, due to the rearrangement shown in Fig. 5b, the half-width of the intra-tube transitions (corresponding to the shifted extrema) increases and its intensity decreases [27].

In order to obtain the unrolled reciprocal lattice of a DWCNT and to check the proximity of the points $\mathbf{K_1}$ and $\mathbf{K'_1}$, one can use the expressions from Appendix A, or superimpose the flat reciprocal spaces of the SWCNTs. For a proper overlap, not only the directions of the chiral vectors $\mathbf{C}$ and $\mathbf{C'}$ of the inner and outer tubes must coincide, but also the angular lengths of these vectors ($2\pi$) must be the same. Hence, in the direction perpendicular to the common axis, it is necessary to uniformly deform the reciprocal space of one of the tubes, so the perimeters $\mathbf{C}$ and $\mathbf{C'}$ of the unrolled nanotubes coincide [27,31,33]. In the resulting geometric structure, the cutting lines [17] of both SWCNTs also coincide.

Let us proceed directly to the method for calculating the energies of ITTs. Again, since the conduction band is replaced by the exciton band [35,36], we cannot directly use the energies given by the Hamiltonian (11) to determine the ITT energies. However, we can calculate the shift $\Delta E_{ITT}$ in the transition energy when a DWCNT is formed from two pristine SWCNTs as

$$\Delta E_{ITT} = E_{DW} - \frac{E_{in} + E_{out}}{2},$$

where $E_{in}$ and $E_{out}$ are the energies of the inner and outer SWCNTs calculated using NN TBA, $E_{DW}$ is the transition energy calculated using the Hamiltonian (11). Furthermore, to obtain the transition energy, it is necessary to sum up $\Delta E_{ITT}$ and the half-sum of the experimental energy values of or the values from the table [37]. Naturally, such an estimate assumes that the valence and conduction bands in SWCNTs are symmetric, and we use this approximation based on the results from Ref. [37].

We now explore the existence of ITTs in individual DWCNTs, grown by catalytic chemical vapour deposition method over open slits (see Ref. [34] for more details on the synthesis procedure). These DWCNTs were index-assigned by analyzing the electron diffraction patterns following the approach [44]. Among tens of probed DWCNTs, we selected two satisfying the condition for ITTs according to our theory, namely (10,6)@(14,13) and (14,8)@(19,14).

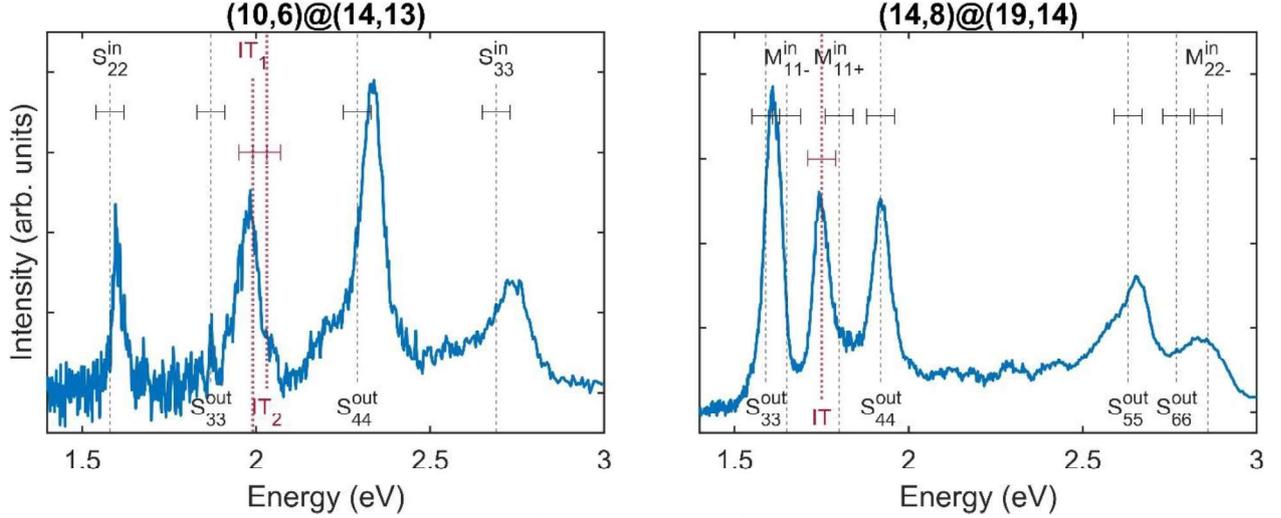

FIG. 6. Rayleigh scattering spectra for DWCNTs (10,6)@(14,13) and (14,8)@(19,14). The dashed vertical lines show the energies of optical transitions predicted by our theory. The labels of the transitions originating from the inner and outer SWCNTs are shown at the top and bottom parts of the panels, respectively. Red dotted lines correspond to the inter-tube transitions.

Fig. 6 presents their Rayleigh scattering spectra (solid blue lines) with the energies of optical transitions calculated within the developed approach. The energies of intra-tube transitions (dashed lines) were found using Eqs. (15-16), while ITT energies (marked as IT in Fig. 6) were calculated using the Hamiltonian (11) and Eq. (16). The horizontal error bars indicate the uncertainty in optical transition energies calculated for pristine SWCNTs. The uncertainty (±40 meV that is two standard deviations) results from using the formula of Liu et al. [37].

We note that there are two ways to interpret the spectra of the DWCNTs in Figure 6. Since the electron diffraction pattern analysis cannot distinguish the handedness of DWCNT layers, we have to consider two possible geometrical configurations, i.e. with the same and opposite handednesses (denoted in the following as (+1) and (-1) DWCNTs, respectively). The latter DWCNTs do not have the ITTs since the points $K_1$ and $K'_1$ in this configuration always have a significant spacing (see detailed geometrical analysis in the next section). Hence, their optical spectra should contain only a superposition of intra-tube transitions (Table 1). On the other hand, DWCNT(+1) satisfy the geometrical conditions for ITTs. Therefore, the peaks at 1.98 and 1.74 eV in (10,6)@(14,13) and (14,8)@(19,14) Rayleigh spectra, respectively, can be a superposition of intra-tube transitions and ITTs. Let us note that Rayleigh spectra of the DWCNT (14,8)@(19,14) in Figure 6 can be fitted using both sets of transitions from Table 1, i.e. with and without ITTs. Nevertheless, considering the accuracy of the theory proposed, the peak at 1.98 eV in (10,6)@(14,13) can hardly be associated with the nearest $S_{33}^{in}$ intra-tube transition, and, in our opinion, the interpretation shown in Fig. 6 is much more likely. In addition, our previous spectroscopic study [34] and independent HRTEM study [45] suggest that DWCNT(+1) configurations are more abundant than DWCNT(-1). In any case, further spectroscopic studies, which go outside the scope of our largely theoretical work, are necessary to establish the origin of the considered resonances exactly.

Let us note that the developed theory also agrees with the results obtained in Ref. [28]. In this article, the author demonstrates the possibility of inter-tube electron transitions on the example of DWCNT (15,13)@(21,17). The theory [28] predicts four ITTs in this tube, specifically, a doublet with energies of 1.34 eV and 1.40 eV and a doublet with energies of 2.26 eV and 2.31 eV. Our model predicts the same pairs of ITTs however, the energies of these doublets calculated using Eq. (11) and Eq. (16) turn out to be a bit different: 1.17 eV, 1.28 eV and 2.23 eV, 2.35 eV

In the conclusion of this section, we would like to emphasize that the developed tight-binding theory has many potential applications. For example, it can be used to calculate conductance, optical absorption or even the magic angles in TBLG. However, in view of the growing amount of experimental data on the optical spectra of DWCNTs [26,33,38,46,47] and the discovery of novel inter-tube electron transitions [26–28], in this work we decided to focus on calculations of optical transition energies, which is also the simplest way to compare our theory with an experiment.

TABLE 1. Energies of optical transitions in DWCNTs (10,6)@(14,13) and (14,8)@(19,14). The SW row shows the energies of pristine SWCNTs comprising corresponding DWCNTs. Their values are taken from Ref. [37]. The DW(+1) and DW(-1) rows correspond to the calculated energies in DWCNTs with the same (+1) and opposite (-1) handednesses.

| | Transition | $S_{22}^{in}$ | $S_{33}^{in}$ | $ITT_1$ | $ITT_2$ | $S_{33}^{out}$ | $S_{44}^{out}$ |
|---|---|---|---|---|---|---|---|
| DWCNT (10,6)@(14,13) | SW, eV | 1.66 | 2.74 | - | - | 1.93 | 2.34 |
| | DW(-1), eV | 1.59 | 2.69 | - | - | 1.88 | 2.28 |
| | DW(+1), eV | 1.58 | 2.69 | 1.99 | 2.03 | 1.87 | 2.29 |

| | Transition | $M_{11-}^{in}$ | $M_{11+}^{in}$ | $M_{22-}^{in}$ | $ITT_1$ | $ITT_2$ | $S_{33}^{out}$ | $S_{44}^{out}$ | $S_{55}^{out}$ | $S_{66}^{out}$ |
|---|---|---|---|---|---|---|---|---|---|---|
| DWCNT (14,8)@(19,14) | SW, eV | 1.69 | 1.80 | 2.91 | - | - | 1.69 | 1.98 | 2.70 | 2.84 |
| | DW(-1), eV | 1.63 | 1.73 | 2.86 | - | - | 1.62 | 1.92 | 2.64 | 2.79 |
| | DW(+1), eV | 1.65 | 1.80 | 2.86 | 1.75 | 1.75 | 1.59 | 1.92 | 2.63 | 2.77 |

## V. DISCUSSION AND CONCLUSION

In this work we develop a theory of the electron band structure in bilayer carbon nanosystems. The proposed method for the construction of the Hamiltonian uses the basis of linearly independent Bloch functions shifted in the reciprocal space. The introduction of this basis is caused by the lowering of the translational symmetry in the considered bilayer structures. Within the developed approach, to construct an effective Hamiltonian one can choose only a few critical functions from the basis and consider the formation of the low-energy band structure from the folded (or shifted) dispersion curves. One of the main advantages of the developed theory is that it allows considering the electron properties of commensurate and incommensurate bilayer structures from a unified point of view. The effective Hamiltonians used to calculate the electron spectra are constructed according to the common principles. The small size of the matrices of these Hamiltonians together with the simplified expressions for the inter-layer coupling provide an efficient way to calculate and analyze the band structure of the investigated systems.

Strictly speaking, the use of the wave vector (and Bloch functions) to characterize the modes of an incommensurate system is not completely correct, and an alternative approach is to consider a very large finite fragment of an infinite structure and diagonalize Hamiltonian matrices with the correspondingly large dimensions [28,48]. However, in the vicinity of Dirac points, the concept of a wave vector turns out to be quite applicable for the lowest-energy electron excitations in the bilayer incommensurate systems under consideration. It is similar to many other aperiodic systems, where low-frequency sound with a wavelength much larger than the size of characteristic structural units can be characterized by a wave vector [49]. The applicability of the wave vector concept, proven for DWCNTs by the comparison with experimental data, can also be justified by the fact that the graphene layers interact via relatively weak Van der Waals forces and the atoms of one layer are located more or less randomly with respect to the atoms of the adjacent layer. The use of the wave vector turns out to be very beneficial also for inter-tube transitions, when one of the bands involved in the transition originates predominantly from one pristine nanotube, and the other predominantly comes from another tube [27].
Continuum models [15,20–22,24] are often used as an alternative approach to calculate the band structure in incommensurate objects. Within this framework, the inter-layer coupling is expanded into series in the moiré vectors $\Delta \boldsymbol{b}_1$ and $\Delta \boldsymbol{b}_2$ of the reciprocal lattice. It is also assumed that Fourier transform of the inter-layer hopping integral $u(\rho, d)$ decays quite fast with increasing distance in the reciprocal space. Thus, this expansion is limited only to a few first Fourier components. Indeed, the inter-layer matrix elements decay according to the exponential law. However, the applicability of such an expansion for structures with only an approximate periodicity is a matter of debates, since the value of the matrix elements in this expansion will depend on the choice of the unit cell. In our opinion, the approach [27,33] is more convincing. Its main assumption is that in an incommensurate bilayer, on average, atoms of different layers are located relative to each other in an approximately random manner. Therefore, to calculate the inter-layer coupling we recommend using Eq. (9), or when the low-energy region of the band structure is considered, one can use the approximate expression Eq. (C6) from Appendix C.

Within the framework of our approach, it is also possible to find more accurate conditions allowing ITTs and show that DWCNTs, where ITTs are possible, are characterized by large unit cells of moiré patterns and a specific relation between the chirality indices. Let us consider the unrolled net of DWCNT. For the sake of clarity, we assume that the sheet of the inner nanotube is stretched $\varepsilon$ times, where $\varepsilon = R_{out}/R_{in}$, and the sheet of the outer tube is not deformed, i.e. $\varepsilon = 1$. Then (in comparison with the expressions from Appendix A), the reciprocal space vectors of the inner and outer nanotubes unrolled into the sheets are redefined as follows

$$\boldsymbol{b}_1 = \left[\frac{2\pi n}{\sqrt{3T}a_0\varepsilon}, \frac{2\pi(2m+n)}{3\sqrt{T}a_0}\right], \quad (17)$$
$$\boldsymbol{b}_2 = \left[\frac{2\pi m}{\sqrt{3T}a_0\varepsilon}, -\frac{2\pi(2n+m)}{3\sqrt{T}a_0}\right],$$

where $T = n^2 + m^2 + nm$. Similar to the TBLG case, using the vectors $\Delta \boldsymbol{b}_1 = \boldsymbol{b}_1^{in} - \boldsymbol{b}_1^{out}$, $\Delta \boldsymbol{b}_2 = \boldsymbol{b}_2^{in} - \boldsymbol{b}_2^{out}$ and orthogonality relations, one can construct a moiré lattice in the direct space. The area $S^M$ of the moiré unit cell can be found as

$$S^M = \frac{(2\pi)^2}{|\Delta \boldsymbol{b}_1 \times \Delta \boldsymbol{b}_2|}. \quad (18)$$

Earlier [27] we formulated the conditions allowing ITTs as a requirement for the proximity of points $\boldsymbol{K}_1$ and $\boldsymbol{K}'_1$. Namely, these two points should not be spaced by more than two cutting lines, while the difference of their $k$-coordinates along the DWCNT axis should also be small. Using Eq. (17), it is easy to show that the first of these conditions is equivalent to the following one

$$|n_{in} - m_{in} - n_{out} + m_{out}| < 6, \quad (19)$$

where $(n_{in}, m_{in})$ and $(n_{out}, m_{out})$ are the chirality indices of the inner and outer tubes, respectively. Using the expressions (17) one can also demonstrate that the smaller the difference of $k$-coordinates of the points $\boldsymbol{K}_1$ and $\boldsymbol{K}'_1$ is, the more accurate the relation below is

$$\frac{n_{out} + m_{out}}{n_{in} + m_{in}} \approx \frac{R_{out}}{R_{in}}. \quad (20)$$

Let us show that satisfaction of the conditions (19-20) for a given DWCNT corresponds to a large unit cell of the moiré pattern in this tube. To do so, consider the problem of maximizing $S^M$ with respect to the indices of the outer and inner tubes. Suppose that $n_{out} = x \cdot n_{in}$ and $m_{out} = y \cdot m_{in}$. If we assume, when formally finding the extremum of $S^M$, that the tube indices are real and not necessarily integers,

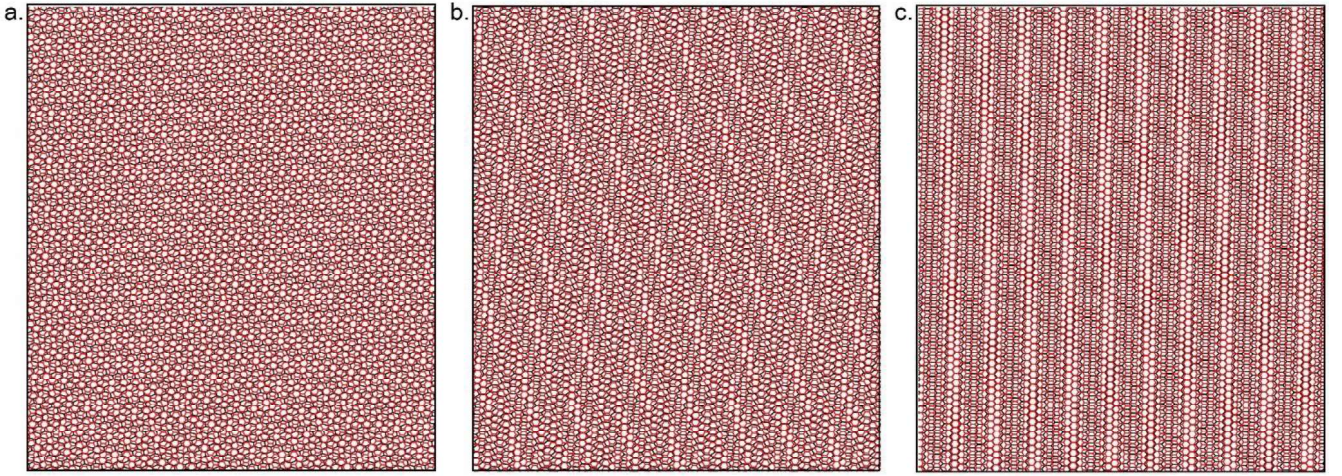

FIG. 7. Moiré patterns in DWCNTs (14,2)@(15,13), (10,6)@(14,13) and (12,11)@(17,16). (a) The unrolled direct space of DWCNT (14,2)@(15,13). In this nanotube, $\Delta\varphi=21.052°$ and ITTs are forbidden according to our theory. One can see an aperiodic motif in the pattern, and the regions of local similarity are very small. (b) The unrolled direct space of DWCNT (10,6)@(14,13). In this tube inter-tube transitions are possible, and the points $K_1$ and $K_1'$ are quite close. The regions of local similarity become elongated in comparison with those in the panel (a). (c) The unrolled direct space of DWCNT (17,16)@(12,11). The chirality vectors of the outer and inner tubes are almost collinear, which corresponds to the extremely elongated moiré pattern. Moreover, the points $K_1$ and $K_1'$ almost coincide in the scheme of superimposed zones.

then the solution of the equation system $S^M/\partial x = 0$ and $\partial S^M/\partial y = 0$ is

$$x = y = \frac{R_{out}}{R_{in}}. \quad (21)$$

Accordingly, the larger the area $S^M$ is, the more accurate the following relations are

$$\frac{n_{out}}{n_{in}} \approx \frac{m_{out}}{m_{in}} \approx \frac{R_{out}}{R_{in}}. \quad (22)$$

Note that Eq. (22) is equivalent to the condition that the difference $\Delta\varphi$ of the chiral angles of the DWCNT layers is close to zero. The relations (22) also become quite accurate if the left-hand side of the inequality (19) vanishes. However, one should not a priori [without direct calculation of Eq. (18)] assume that a DWCNT, in which the left-hand side of Eq. (19) is zero, always has a larger value of $S^M$ than another DWCNT whose chiral indices satisfy Eq. (20) with better accuracy.

For the DWCNT (12,11)@(17,16) (in which ITTs were detected experimentally for the first time [26]) the right-hand side of Eq. (19) is equal to 0. In this DWCNT $\Delta\varphi = 0.436°$ and $S^M = 1532.06$ nm$^2$. Among the nanotubes that we discussed in this work, and in which we suppose the presence of ITTs, the largest value of $\Delta\varphi=6.988°$ corresponds to the DWCNT (10,6)@(14,13). The area of the moiré cell in this tube is $S^M=7.37$ nm$^2$ and is hundreds of times smaller than in the DWCNT (12,11)@(17,16) (see the moiré patterns in above DWCNTs in Fig. 7). Thus, the analysis performed shows that simply large (not necessarily extremely large) areas $S^M$ and, accordingly, small angles $\Delta\varphi$ correspond to DWCNTs with strong inter-layer coupling and possible inter-tube transitions.

In all the considered DWCNTs where ITTs can occur, the values of $\Delta\varphi$ are small and the points $K_1$ and $K_1'$ are close. The latter brings to a natural question: are there such incommensurate DWCNTs in which the points $K_2$ and $K_2'$ or $K_3$ and $K_3'$ are close instead of the points $K_1$ and $K_1'$? Using Eq. (17), it is easy to obtain the conditions for the proximity of these points and then carry out the corresponding analysis, for example, using a simple computer program. It turns out that in the extremely rare cases of DWCNTs selected this way, the points $K_1$ and $K_1'$ are always going to be even closer. Thus, the pairs of $K$-points in DWCNTs are fundamentally nonequivalent, and if there is indeed a strong coupling between the DWCNT layers, then it is due to the proximity of the first pair of points.

In conclusion, in this work we have developed a theory of the band structure in TBLG and incommensurate DWCNTs. In particular, we found more accurate conditions which allow ITTs and demonstrated ITT presence in the DWCNTs (10,6)@(14,13). Using the simplified method for calculating the inter-layer matrix elements in incommensurate bilayer carbon nanostructures, we calculated the energies of intra- and inter-tube optical transitions in 30 DWCNTs. The proposed theory is in excellent agreement with all known experimental data on optical transitions [26,33,46,47]. This theory can be easily generalized to the case of other incommensurate bilayer systems where the periodic moiré patterns emerge [50–55], and also can be used to calculate the band structure in promising trilayer graphene structures, where, as in TBLG, novel Mott insulating and superconducting states were discovered [56–58].

## ACKNOWLEDGEMENTS

The authors thank Matthieu Paillet for the help in measuring Rayleigh spectra of individual DWCNTs. D.Ch., D.L. and S.R. acknowledge financial support from the Russian Foundation for Basic Research (Grant No. 18-29-19043 mk).

## APPENDIX A. DESCRIPTION OF THE DIRECT AND RECIPROCAL SPACES OF SWCNT

In regard to SWCNT translational and rotational symmetry, it is convenient to express the graphene basis translations $\boldsymbol{a}_1$ and $\boldsymbol{a}_2$ in the cylindrical coordinate system.

By simply projecting these vectors along the perimeter and the axis of a SWCNT with the chiral indices $(n, m)$, we obtain the first and the second components of the SWCNT basis translations, respectively. In the explicit form they are written as [27]:

$$\boldsymbol{a}_1 = \left[\frac{(2n+m)\pi}{(n^2+m^2+nm)}, \frac{3ma_0}{2\sqrt{n^2+m^2+nm}}\right], \quad (A1)$$

$$\boldsymbol{a}_2 = \left[\frac{(2m+n)\pi}{(n^2+m^2+nm)}, \frac{-3na_0}{2\sqrt{n^2+m^2+nm}}\right], \quad (A2)$$

where $a_0 = 0.142$ nm is the length of the carbon bond. Using the relations $\boldsymbol{a}_i \boldsymbol{b}_j = 2\pi \delta_{ij}$, one can easily find the components of the reciprocal lattice vectors $\boldsymbol{b}_1$ and $\boldsymbol{b}_2$:

$$\boldsymbol{b}_1 = \left[n, \frac{2\pi}{3a_0}\frac{2m+n}{\sqrt{n^2+m^2+nm}}\right], \quad (A3)$$

$$\boldsymbol{b}_2 = \left[m, -\frac{2\pi}{3a_0}\frac{2n+m}{\sqrt{n^2+m^2+nm}}\right]. \quad (A4)$$

Within the proposed definition of basis translations, the first component of the vectors (A3-A4) becomes dimensionless since it measures the distance in units equal to the distance between the nearest cutting lines. Thus, the cutting lines of inner and outer tubes forming a DWCNT will automatically match in the scheme of superimposed extended zones.

## APPENDIX B. EQVIVALENT APPROACH TO CONSTRUCT TBLG HAMILTONIAN

In order to calculate the TBLG band structure one can use a different set of wave functions instead of Eq. (4) and correspondingly a different Hamiltonian. Let us consider the following set

$$\varphi_k(\boldsymbol{r}) = \sqrt{\frac{T}{N}} \sum_{\boldsymbol{R}_k} \phi(\boldsymbol{r} - \boldsymbol{R}_k) e^{I\boldsymbol{q}\cdot\boldsymbol{R}_k}, \quad (B1)$$

where the integer $k = 1 \ldots 4T$ indexes the sub-sublattices, $\phi(\boldsymbol{r})$ is the atomic $p_z$-orbital, $\boldsymbol{R}_k$ is the coordinate of an atom belonging to the sub-sublattice $k$, $N$ is the number of hexagons in the top (or bottom) layer of graphene. Within the proposed basis the Hamiltonian matrix elements are written as

$$H'^b_{i,j} = \frac{T}{N}\sum_{\boldsymbol{R}_i, \boldsymbol{R}_j} e^{I\boldsymbol{q}\cdot(\boldsymbol{R}_j - \boldsymbol{R}_i)} u(\boldsymbol{R}_i, \boldsymbol{R}_j), \quad (B2)$$

where the coordinates $\boldsymbol{R}_i$ and $\boldsymbol{R}_j$ correspond to different sub-sublattices. As before, when calculating the sum over the sub-sublattice $i$, we can consider only one atom in this sub-sublattice. Then Eq. (B2) is rewritten as

$$H'^b_{i,j} = \sum_{\boldsymbol{R}_j} e^{I\boldsymbol{q}\cdot(\boldsymbol{R}_j - \boldsymbol{R}_i)} u(\boldsymbol{R}_i, \boldsymbol{R}_j). \quad (B3)$$

The band structures calculated within the Hamiltonians (B3) and (5) numerically coincide.

## APPENDIX C. INTER-LAYER MATRIX ELEMENTS IN INCOMMENSURATE BILAYER SYSTEMS

Let us consider the coupling between the modes with wave vectors $\boldsymbol{q}$ and $\boldsymbol{q} - \boldsymbol{Q}_s$ of the top and bottom graphene layers, respectively. Let $\boldsymbol{R}$ be the coordinate of a node in the sublattice $i$ of the top layer, and $\boldsymbol{R}'$ be the coordinate of a node in the sublattice $j$ of the bottom layer. Then the inter-layer coupling matrix element is written as:

$$V(\boldsymbol{q}, \boldsymbol{Q}_s) = \frac{1}{N}\sum_{\boldsymbol{R}, \boldsymbol{R}'} e^{I[\boldsymbol{q}\cdot\boldsymbol{R} - (\boldsymbol{q}-\boldsymbol{Q}_s)\cdot\boldsymbol{R}']} u(|\boldsymbol{\rho}|, d), \quad (C1)$$

where $N$ is the number of hexagons in the top (or bottom) monolayer of graphene, $d = 0.34$ nm is the inter-layer distance and $\boldsymbol{\rho} = \boldsymbol{R} - \boldsymbol{R}'$. Let us rewrite Eq. (C1) as:

$$V(\boldsymbol{q}, \boldsymbol{Q}_s) = \frac{1}{N}\sum_{\boldsymbol{R}} e^{I\boldsymbol{Q}_s\cdot\boldsymbol{R}} \\ \times \sum_{\boldsymbol{R}'} e^{I(\boldsymbol{q}-\boldsymbol{Q}_s)\cdot(\boldsymbol{R}-\boldsymbol{R}')} u(|\boldsymbol{\rho}|, d), \quad (C2)$$

Following Refs. [27,33], we consider the relative positions of atoms in different layers of incommensurate TBLG approximately random. Then the in-plane distance $\boldsymbol{R} - \boldsymbol{R}'$ runs over an almost continuous set of values and we can replace the sum by an integral

$$V(\boldsymbol{q}, \boldsymbol{Q}_s) = \frac{1}{NS_0}\sum_{\boldsymbol{R}} e^{I\boldsymbol{Q}_s\cdot\boldsymbol{R}} \\ \times \int_{S'} e^{I(\boldsymbol{q}-\boldsymbol{Q}_s)\cdot(\boldsymbol{R}-\boldsymbol{R}')} u(|\boldsymbol{\rho}|, d) dS', \quad (C3)$$

where $S_0$ is the area of the graphene unit cell and $S'$ is the bilayer graphene area. It is easy to see that the integral in Eq. (C3) is simply the Fourier transform of the function $u(|\boldsymbol{\rho}|, d)$, as which it is convenient to choose a Gaussian one

$$u(|\boldsymbol{\rho}|, d) = \gamma'_c \exp[-\alpha(\rho^2 + d^2)]. \quad (C4)$$

Let us consider separately the integral from Eq. (C3):

$$\frac{\gamma'_c}{S_0} e^{-\alpha d^2} \int_{S'} e^{I(\boldsymbol{q}-\boldsymbol{Q}_s)\cdot\boldsymbol{\rho}} e^{-\alpha \rho^2} dS'. \quad (C5)$$

Substituting $\boldsymbol{\rho} = (x, y)$ and integrating over the infinite area of the bilayer, we reduce Eq. (C5) to the following expression:

$$\frac{\gamma'_c \pi}{S_0 \alpha} \exp(-\alpha d^2) \exp\left[-\frac{(\boldsymbol{q}-\boldsymbol{Q}_s)^2}{4\alpha}\right].$$

Thus, the inter-layer coupling matrix elements (C1) decay exponentially with an increase in the distance $|\boldsymbol{q} - \boldsymbol{Q}_s|$. Further, the factor $\frac{1}{N}\sum_{\boldsymbol{R}} \exp(I\boldsymbol{Q}_s \cdot \boldsymbol{R})$ in Eq. (C3) is equal to 1 if $\boldsymbol{Q}_s = 0$. The modulus of this factor is also equal to 1, if $\boldsymbol{Q}_s$ is a translation vector of the *top* layer reciprocal lattice. In this case (when choosing the origin at the center of a carbon hexagon) the phase of this expression can take only the following three values: $0, \pm 2\pi/3$. In all the other cases, Eq. (C1) vanishes, and there is simply no coupling between the modes $\boldsymbol{q}$ and $\boldsymbol{q} - \boldsymbol{Q}_s$. Let us note, since the twisted bilayer graphene is symmetric with respect to layer permutations, then there also should be coupling between the modes, whose wave vectors differ by any vector $\boldsymbol{Q}'_s$ of the *bottom* layer reciprocal lattice and the distance $|\boldsymbol{q} - \boldsymbol{Q}_s'|$ is relatively small. Thus, in the cases when $\boldsymbol{Q}_s = 0$ the matrix elements can be approximately calculated as

$$v(\boldsymbol{q}) = \frac{\gamma'_c \pi}{S_0 \alpha} \exp(-\alpha d^2) \exp\left(-\frac{q^2}{4\alpha}\right). \quad (C6)$$

If $\boldsymbol{Q}_s$ is expressed as an integer linear combination of basis vectors of the top (or bottom) reciprocal lattice, the matrix elements are found according to Eq. (10). Otherwise, $V(\boldsymbol{q}, \boldsymbol{Q}_s) = 0$.

The interlayer matrix elements in a DWCNT can be calculated in a similar way. To do so, we carry out the transition to the cylindrical coordinate system and note that the function (C4) depends only on the distance $l$ between the points $\boldsymbol{R}$ and $\boldsymbol{R}'$ on the surface of two cylinders: $u(\Delta\varphi, \Delta z) \equiv u(l)$. Then

$$u(l) \approx \gamma_c' e^{-\alpha(R_{in}R_{out}\Delta\varphi^2 + \Delta z^2 + \Delta R^2)}, \quad (C7)$$

where $R_{in}$ and $R_{out}$ are radii of the inner and outer SWCNT, $\Delta R = R_{out} - R_{in}$ is the inter-layer distance in the DWCNT, $\Delta\varphi = \varphi - \varphi'$, $\Delta z = z - z'$. Deriving Eq. (C7), we used the expansion of $\cos\Delta\varphi$, which is justified by the fact that in concentric nanotubes two orbitals with an angular spacing practically do not overlap. Let us consider first the case $\boldsymbol{Q}_s = 0$ and rewrite (C2) as

$$v(\boldsymbol{q}) = \frac{1}{\sqrt{N_{in}N_{out}}} \sum_{\boldsymbol{R'}} \sum_{\boldsymbol{R}} e^{i\boldsymbol{q}\cdot(\boldsymbol{R}-\boldsymbol{R'})} u(l)$$
$$\approx \sum_{\boldsymbol{R'}} \int_S \frac{\cos(\mu\Delta\varphi)\cos(k\Delta z) u(\Delta\varphi, \Delta z) dS}{S_0\sqrt{N_{in}N_{out}}}, \quad (C8)$$

where $\mu$ and $k$ are the wave vector $\boldsymbol{q}$ components in the SWCNT, $S$ is the area of the inner tube, $dS = R_{in}d\varphi dz$, $N_{in}$ and $N_{out}$ are the hexagon numbers in the inner and outer layers of DWCNT, respectively. In Eq. (C8), similar to the flat case, we replace the summation over $\boldsymbol{R}$ by an integration over the area of the inner nanotube. Since the function $u(\Delta\varphi, \Delta z)$ is even with respect to the arguments $\Delta\varphi$ and $\Delta z$, Eq. (C8) is also real. Making the substitution $\Delta\varphi \to \varphi$, $\Delta z \to z$ and performing the summation over the nodes of the outer tube $\boldsymbol{R'}$, we obtain the following integral:

$$v(\boldsymbol{q}) = C \int_{-\pi}^{\pi} \cos(\mu\varphi) e^{-\alpha R_{in}R_{out}\varphi^2} d\varphi$$
$$\times \int_{-\infty}^{+\infty} \cos(kz) e^{-\alpha z^2} dz, \quad (C9)$$

where $C = \gamma_c' \exp(-\alpha \Delta R^2) \sqrt{R_{out}R_{in}}/S_0$. Deriving Eq. (C9) we also used the relation $\sqrt{N_{out}/N_{in}} = \sqrt{R_{out}/R_{in}}$. Due to the fast convergence of the function $\exp(-\alpha R_{in}R_{out}\varphi^2)$, we can extend the integration limits and integrate from $-\infty$ to $+\infty$. As a result, we obtain

$$v(\mu, k) = \frac{\gamma_c'\pi}{S_0\alpha} \exp(-\alpha \Delta R^2) \exp\left(-\frac{\boldsymbol{q}_{in}\cdot\boldsymbol{q}_{out}}{4\alpha}\right) \quad (C10)$$

where $\boldsymbol{q}_{in}\cdot\boldsymbol{q}_{out} = k^2 + \mu^2/(R_{in}R_{out})$ is a scalar product of the wave vectors in the inner and outer tubes. The inter-layer matrix elements for $\boldsymbol{Q}_s \neq 0$ are found according to the same principle as in the case of bilayer graphene. If $\boldsymbol{Q}_s$ is a translation of the inner (or outer) tube reciprocal lattice, one should use Eq. (10). Otherwise, $V(\boldsymbol{q}, \boldsymbol{Q}_s) = 0$.

Below, for reference, we give expressions for the inter-layer matrix elements obtained as the Fourier transform of the function $u(\rho, d) = \gamma_c \exp[-(\sqrt{\rho^2 + d^2} - d)/\lambda]$ (in the case of TBLG) and the function $u(l) = \gamma_c \exp[-(l-d)/\lambda]$ (in the case of DWCNT):

$$v^{TBLG}(\boldsymbol{q}) = \frac{4\gamma_c}{\lambda S_0} \exp\left(\frac{d}{\lambda}\right) \sqrt{\frac{\pi d^3}{2}} \left(q^2 + \frac{1}{\lambda^2}\right)^{-\frac{3}{4}} K_{-\frac{3}{2}}\left[d\left(q^2 + \frac{1}{\lambda^2}\right)^{1/2}\right], \quad (C11a)$$

$$v^{CNT}(\mu, k) = \frac{4\gamma_c}{\lambda S_0} \exp\left(\frac{d}{\lambda}\right) \sqrt{\frac{\pi \Delta R^3}{2}} \left(\boldsymbol{q}_{in}\boldsymbol{q}_{out} + \frac{1}{\lambda^2}\right)^{-\frac{3}{4}} K_{-\frac{3}{2}}\left[\Delta R\left(\boldsymbol{q}_{in}\cdot\boldsymbol{q}_{out} + \frac{1}{\lambda^2}\right)^{1/2}\right], \quad (C11b)$$

where $K_\nu(x) = \int_0^\infty \exp[-x\cosh(t)]\cosh(\nu t)\,dt$ is modified Bessel function of the second kind.

Expanding into series Eqs. (C11a, C11b) in $\lambda/d$ (or $\lambda/\Delta R$), one can obtain the following simplified expressions for the matrix elements:

$$v^{TBLG}(\boldsymbol{q}) = \frac{2\pi\lambda d}{S_0}\gamma_c e^{-\lambda d \frac{q^2}{2}}, \quad (C12a)$$

$$v^{CNT}(\mu, k) = \frac{2\pi\lambda\Delta R}{S_0}\gamma_c e^{-\frac{\Delta R - d}{\lambda}} e^{-\lambda\Delta R\frac{\boldsymbol{q}_{in}\cdot\boldsymbol{q}_{out}}{2}}. \quad (C12b)$$

Note that the matrix elements (C12a, C12b) also decay according to the same exponential law as the functions (C6) and (C10), respectively. Eq. (C12b) coincides with the expression for the inter-layer matrix elements from Ref. [33] up to a constant factor.

---